\documentclass[12pt]{article}
\usepackage{amssymb}
\usepackage{amsopn}
\def\hybrid{\topmargin 0pt	\oddsidemargin 0pt 
	\headheight 0pt	\headsep 0pt
	\textheight 9in		
	\textwidth 6.1in	
	\marginparwidth .875in
	\parskip 5pt plus 1pt	\jot = 1.5ex}

\hybrid

\catcode`\@=11

\def\numberbysection{\@addtoreset{equation}{section}
	\def\theequation{\thesection.\arabic{equation}}}

\def\underline#1{\relax\ifmmode\@@underline#1\else
	$\@@underline{\hbox{#1}}$\relax\fi}

\def\titlepage{\@restonecolfalse\if@twocolumn\@restonecoltrue\onecolumn
     \else \newpage \fi \thispagestyle{empty}\c@page\z@	
	\def\thefootnote{\fnsymbol{footnote}} }

\def\endtitlepage{\if@restonecol\twocolumn \else \newpage \fi
	\def\thefootnote{\arabic{footnote}} 
	\setcounter{footnote}{0}}  

\catcode`@=12
\relax
\numberbysection
\DeclareMathOperator{\so}{\mathfrak{s}\mathfrak{o}}
\DeclareMathOperator{\Tr}{Tr}
\DeclareMathOperator{\Orth}{O}
\DeclareMathOperator{\SOrth}{SO}
\DeclareMathOperator{\SL}{SL}
\DeclareMathOperator{\Ad}{Ad}
\DeclareMathOperator{\ad}{ad}
\newcommand{\hodge}{*_{\Sigma}}
\newcommand{\lieg}{\mathfrak{g}}

\newcommand{\lieh}{\mathfrak{h}}
\newcommand{\liem}{\mathfrak{m}}

\newcommand{\Btil}{\widetilde{B}}
\newcommand{\Gtil}{\widetilde{G}}
\newcommand{\Htil}{\widetilde{H}}
\newcommand{\Qtil}{\widetilde{Q}}
\newcommand{\Mtil}{\widetilde{M}}
\newcommand{\Rtil}{\widetilde{R}}
\newcommand{\Tij}{T^{i}{}_{j}}
\newcommand{\Tik}{T^{i}{}_{k}}
\newcommand{\Tjk}{T^{j}{}_{k}}
\newcommand{\Tkj}{T^{k}{}_{j}}
\newcommand{\Ttil}{\widetilde{T}}
\newcommand{\alphatil}{\tilde{\alpha}}
\newcommand{\dalem}{\partial^{2}_{+-}}
\newcommand{\dminus}{\partial_{-}}
\newcommand{\dplus}{\partial_{+}}
\newcommand{\gtil}{\tilde{g}}
\newcommand{\half}{\frac{1}{2}}
\newcommand{\ktil}{\tilde{k}}
\newcommand{\bbR}{\mathbb{R}}
\newcommand{\omegatil}{\tilde{\omega}}
\newcommand{\sigtil}{\tilde{\sigma}}
\newcommand{\rtil}{\tilde{r}}
\newcommand{\thetatil}{\tilde{\theta}}
\newcommand{\xtil}{\tilde{x}}

\begin{document}
\begin{titlepage}
\mbox{April 2000}\hfill\mbox{UMTG--223}\newline
\strut\hfill\mbox{\tt hep-th/0004120}
\par\vspace{.5in}
\begin{center}
    {\bf\large Target Space Pseudoduality}\\
    {\bf\large Between Dual Symmetric Spaces\footnote{This work
    was supported in part by National Science Foundation grant
    PHY--9870101.}}\\[.4in]
    {\bf Orlando Alvarez\footnote{email: 
    \texttt{oalvarez@miami.edu}}}\\
    {\em Department of Physics}\\
    {\em University of Miami}\\
    {\em P.O. Box 248046}\\
    {\em Coral Gables, FL 33124 USA}\\
\end{center}

\vspace{.5in}
\begin{abstract}
A  set of on shell duality equations is proposed that leads to a
map between strings moving on symmetric spaces with opposite
curvatures.  The transformation maps ``waves'' on a riemannian
symmetric space $M$ to ``waves'' on its dual riemannian symmetric
space $\Mtil$.  This transformation preserves the energy momentum
tensor though it is not a canonical transformation.  The preservation
of the energy momentum tensor has a natural geometrical
interpretation. The transformation maps ``particle-like solutions'' into 
static ``soliton-like solutions''. The results presented here 
generalize earlier results of E.~Ivanov.
\end{abstract}

\vspace{.5in}
\noindent
PACS: 11.25-w, 03.50-z, 02.40-k\newline
Keywords: duality, strings, geometry

\end{titlepage}

\section{Introduction}
\label{sec:intro}

We take spacetime $\Sigma$ to be two dimensional Minkowski space.  Let
$\sigma^{\pm} = \tau\pm\sigma$ be the standard lightcone coordinates
and let $\varphi$ be a massless free scalar field satisfying the wave
equation $\dalem\varphi=0$.  The standard abelian duality
transformations (for a review see \cite{Giveon:1994fu}) are
\begin{eqnarray}
    \dplus\tilde{\varphi} & = & +\dplus\varphi\;,
    \label{eq:olddualp}  \\
     \dminus\tilde{\varphi} & = & -\dminus\varphi\;.
    \label{eq:olddualm}
\end{eqnarray}
The integrability conditions for the above are precisely the equations
of motion $\dalem\varphi=0$.  This means that we can construct
$\tilde{\varphi}(\sigma^{+},\sigma^{-})$ and verify that as a
consequence of the above $\dalem\tilde{\varphi}=0$.  A generalization
of the above are the pseudochiral models of the type introduced by
Zakharov and Mikhailov \cite{Zakharov:1978pp}.  
Consider a standard sigma model with target space a Lie group $G$
and with equations of motions $\partial^{a}(g^{-1}\partial_{a}g)=0$. 
Introduce the dual Lie algebra valued field $\phi$ by
$$
    g^{-1}\partial_{a}g = -\epsilon_{a}{}^{b}\partial_{b}\phi\,.
$$
It was shown by Nappi that these two descriptions were not quantum
mechanically equivalent \cite{Nappi:1980ig}.  The correct dual models
were found by Fridling and Jevicki \cite{Fridling:1984ha} and by
Fradkin and Tseytlin \cite{Fradkin:1985ai}.  It was eventually
understood that the duality transformation should be a canonical
transformation \cite{Curtright:1994be,Alvarez:1994wj} and the
transformation in the pseudochiral model is not.  Nevertheless the
pseudochiral model has a variety of interesting field theoretic
features \cite{Curtright:1994be,Zachos:1994fa}.  Motivated by string
theory, there is now a vast literature on nonabelian duality
\cite{Kiritsis:1991zt,Rocek:1992ps,Giveon:1992jj,delaOssa:1993vc,%
Gasperini:1993nz,Giveon:1994mw,Giveon:1994ai,Giveon:1994ph,%
Alvarez:1994zr,Alvarez:1994qi,Lozano:1995jx,Lozano:1996sc,%
Alvarez:1995uc}
and Poisson-Lie duality
\cite{Klimcik:1995ux,Klimcik:1996dy,Klimcik:1996kw,%
Sfetsos:1998pi,Sfetsos:1996xj,Stern:1998my}.

Following up on recent work \cite{Alvarez:2000bh} we consider a
variant of the duality equations proposed there.  These equations are
a generalization of the ones of Zakharov and Mikhailov.  In light of
the introductory paragraph it will be interesting to study the
physical and mathematical properties of these equations.  Here we
point out that there is an interesting transformation that maps
solutions of the wave equation on a symmetric space into solutions of
the wave equation on a symmetric space with the opposite curvature 
(see the next paragraph). 
We note that it was pointed out in \cite{Zachos:1994fa} that the
opposite signs found by Nappi in the beta functions for the dual
models of Zakharov and Mikhailov are explained by observing that the
generalized curvatures in the models have opposite signs.

The results presented here are a generalization\footnote{I am thankful
to E.~Ivanov for bringing his work to my attention.} of results of
E.~Ivanov~\cite{Ivanov:1987yv}.  There is a large body of literature
discussing conserved currents in sigma models based on groups or coset
spaces.  A seminal work was Pohlmeyer's
construction~\cite{Pohlmeyer:1975nb} for an infinite number of
conserved currents in sigma models with target space $S^{n}$.  This
was generalized by Eichenherr and Forger
\cite{Eichenherr:1979ci,Eichenherr:1981sk} who showed that the
construction generalized to symmetric spaces.  Ivanov expanded on
these ideas and in doing so introduced the notion the \emph{dual
algebra} and the \emph{dual sigma model}.  He states, 
\begin{quote}
    ``\ldots\ in
    which we show that the equations of any $d=2$ $\sigma$ model on a
    symmetric space simultaneously describe a $d=2$ $\sigma$ model on some
    other, dual factor space\ldots'' \cite[p.  475]{Ivanov:1987yv}.
\end{quote}
He explicitly works out the example of a sigma model with target a
compact real Lie group $G$ (with zero $B$-field).  He views $G$ as a
symmetric space $G\times G/G$ and he explicitly shows that the sigma
model on $G$ is dual to the sigma model on $G^{\mathbb{C}}/G$ where
$G^{\mathbb{C}}$ is the complexification of $G$.  He subsequently
asserts that the construction generalizes to a generic symmetric 
space. Ivanov's construction for a symmetric space is given in 
Section~\ref{sec:ivanov}.  The work here generalizes Ivanov's in that
we assume a general riemannian manifold and show that it must be a
symmetric space.

We first establish some
notation.  The sigma model with target space $M$, metric $g$ and
$2$-form $B$ will be denoted by $(M,g,B)$ and has lagrangian
\begin{equation}
    \mathcal{L} = \half g_{ij}(x)
    \left(
    \frac{\partial x^{i}}{\partial \tau}
    \frac{\partial x^{j}}{\partial \tau}
    -\frac{\partial x^{i}}{\partial \sigma}
    \frac{\partial x^{j}}{\partial \sigma}
    \right) +
    B_{ij}(x)
    \frac{\partial x^{i}}{\partial \tau}
    \frac{\partial x^{j}}{\partial \sigma}
    \label{eq:lag}
\end{equation}
with canonical momentum density
\begin{equation}
    \pi_{i}= \frac{\partial\mathcal{L}}{\partial \dot{x}^{i}} =
    g_{ij}\dot{x}^{j} + B_{ij}x'^{j}\;.
    \label{eq:canmomentum}
\end{equation}
The stress energy tensor for this sigma model is given by
$\Theta_{+-}=0$,
\begin{equation}
    \Theta_{++}= g_{ij}(x) \dplus x^{i} \dplus x^{j} \quad\mbox{and}\quad
    \Theta_{--}= g_{ij}(x) \dminus x^{i} \dminus x^{j}\,.
    \label{eq:Tmunu}
\end{equation}

In general a duality transformation between sigma models $(M,g,B)$ and
$(\Mtil,\gtil,\Btil)$ is a canonical transformation between the
respective phase spaces that preserves the respective hamiltonian
densities.  We can study a less restrictive situation where we only
have ``on shell duality''.  By this we mean that we only require that
a map exists between solutions to the equations of motion of $(M,g,B)$
and solutions of the equations of motion of $(\Mtil,\gtil,\Btil)$. 
The ``on shell'' transformation proposed below will be
referred to as \emph{pseudoduality}.  There is a certain naturalness
to the pseudoduality transformation because it preserves the stress
energy tensor.

As in \cite{Alvarez:2000bh} it is convenient to choose an orthonormal
frame $\{\omega^{i}\}$ with the antisymmetric riemannian connection
$\omega_{ij}$.  The Cartan structural equations are
\begin{eqnarray*}
    d\omega^{i} & = & -\omega_{ij}\wedge\omega^{j}\;, \\
    d\omega_{ij} & = & -\omega_{ik}\wedge\omega_{kj}
    +\half R_{ijkl}\omega^{k}\wedge\omega^{l}\;.
\end{eqnarray*}
We also define the curvature $2$-forms by $\Omega_{ij} = \half
R_{ijkl}\omega^{k}\wedge\omega^{l}$.  Consider two sigma models
$(M,g,B)$ and $(\Mtil,\gtil,\Btil)$.  Using the orthonormal frame we
define the appropriate $\sigma^{a}$ derivative of the maps
$x:\Sigma\to M$ and $\xtil:\Sigma \to \Mtil$ by\footnote{More
correctly these equations are pullbacks of the type $x^{*}\omega^{i}=
x^{i}{}_{a} d\sigma^{a}$.  In the field of exterior differential
systems \cite{BCG3} the pullback is implicit and not usually written. 
We adhere to that convention.}
\begin{equation}
    \omega^{i}=x^{i}{}_{a}d\sigma^{a}\quad\mbox{and}\quad
    \omegatil^{i}=\xtil^{i}{}_{a}d\sigma^{a}\,.
    \label{eq:defvel}
\end{equation}
In order to describe the equations of motion for the sigma model we
need to consider second derivatives.  The covariant derivatives of
$x^{i}{}_{a}$ are $x^{i}{}_{ab}$ and are defined by
\begin{equation}
    dx^{i}{}_{a} + \omega_{ij}x^{j}{}_{a} = x^{i}{}_{ab}d\sigma^{b}\;.
    \label{eq:covder}
\end{equation}
Taking the exterior derivative of $\omega^{i}=x^{i}{}_{a}d\sigma^{a}$
we learn that $x^{i}{}_{ab}= x^{i}{}_{ba}$.  The equations of motion
coming from (\ref{eq:lag}) are
\begin{equation}
    x^{k}{}_{+-} = -\half H_{kij}x^{i}{}_{+} x^{j}{}_{-}\,,
    \label{eq:eomgen}
\end{equation}
where $H=dB$.  There are similar definitions and equations for
$\xtil^{i}$.

The lagrangian version of the hamiltonian duality equations in
\cite{Alvarez:2000bh} may be written as
\begin{eqnarray}
    \xtil_{+}(\sigma) & = & +T_{+}(x,\xtil) x_{+}(\sigma)\,,
    \label{eq:plus}  \\
    \xtil_{-}(\sigma) & = & -T_{-}(x,\xtil) x_{-}(\sigma)\,.
    \label{eq:minus}
\end{eqnarray}
The orthogonal matrix valued functions $T_{\pm}:M \times \Mtil \to
\SOrth(n)$ are not arbitrary but related by
\begin{equation}
    T_{+}(I+n) = T_{-}(I-n)\,,
    \label{eq:Trel}
\end{equation}
where the antisymmetric tensor $n_{ij}$ on $M\times\Mtil$ satisfies
some PDEs given in \cite{Alvarez:2000bh}.  In this article we relax
such restrictions on $T_{\pm}$ and consider orthogonal matrix valued
functions $T_{\pm}:\Sigma \to \SOrth(n)$ with the constraint that
solutions to the the sigma model $(M,g,B)$ are mapped into solutions
of $(\Mtil,\gtil,\Btil)$ and vice versa.  We will refer to these
equations as the \emph{pseudoduality equations}.  Note that the
pseudoduality transformations satisfy $\widetilde{\Theta}_{\pm\pm}
=\Theta_{\pm\pm}$.  This is very different than in
\cite{Alvarez:2000bh} where $T_{\pm}$ are functions on $M\times
\Mtil$.  In other words, we allow explicit dependence on $\sigma$ in
$T$ where as in \cite{Alvarez:2000bh} the dependence occurs because
$x$ and $\xtil$ depend on $\sigma$.

We specialize in this article to the case where $T_{+}=T_{-}=T: \Sigma
\to \SOrth(n)$ and $H=0$, $\Htil=0$.  In this case the equations of
motion become\footnote{We remind the reader that solutions to
Laplace's equation are called harmonic functions yet solutions to the
wave equation are simply called ``waves''.} $x_{+-}=\xtil_{+-}=0$ and
the duality equations are
\begin{equation}
    \xtil_{\pm}(\sigma) = \pm T(\sigma) x_{\pm}(\sigma)\,.
    \label{eq:special}
\end{equation}
We point out that when $H=\Htil=0$ the sigma models are worldsheet
parity and worldsheet time reversal invariant.  The isometry $T$ is
odd under worldsheet parity and under worldsheet time reversal.  Note
that in the pseudochiral model $\Htil \neq 0$ and therefore it is not
covered in this paper.  The general case with generic $H$ and $\Htil$
will be treated elsewhere~\cite{Alvarez:psd2}.

Duality equations (\ref{eq:special}) are mathematically quite
interesting.  We digress a bit and discuss the question of local
riemannian isometries.  Assume we have a map $f:M\to\Mtil$ and we wish
for this map to be an isometry.  If $\{\omega^{j}\}$ and
$\{\omegatil^{i}\}$ are local orthonormal frames then we require that
the metric be preserved: $f^{*}(\omegatil^{i}\otimes\omegatil^{i}) =
\omega^{j}\otimes\omega^{j}$.  The solution to this equation is
$f^{*}\omegatil = T\omega$ where $T:M \to \Orth(n)$.  This Pfaffian
system of equations is integrable \cite{BCG3} if the Riemann curvature
tensor and its higher order covariant derivatives in the two spaces
agree when identified via $T$.  For more details see the discussion in
the paragraph following (\ref{eq:kktil}).  In our case we begin with
maps $x:\Sigma \to M$, $\xtil:\Sigma \to \Mtil$ that satisfy
hyperbolic Lorentz invariant equations.  We fix a Lorentz frame and we
observe that we are interested in maps from $\{x:\Sigma\to M\,|\,
x_{+-}=0\}$ into $\{\xtil:\Sigma\to\Mtil\,|\, \xtil_{+-}=0\}$ that
preserve the two independent components of the energy momentum tensor
$\Theta_{++}$ and $\Theta_{--}$.  A linearity assumption and
(\ref{eq:Tmunu}) tells us that the maps have to be of form
(\ref{eq:plus}) and (\ref{eq:minus}) with the more general
$T_{\pm}(\sigma)$.  Given a map $x:\Sigma\to M$ there are two
preferred tangent vector fields $\partial/\partial\sigma^{+}$ and
$\partial/\partial\sigma^{-}$.  The conditions that require $\Theta$
to be preserved are very geometric: the lengths of
$\partial/\partial\sigma^{+}$ and $\partial/\partial\sigma^{-}$ are
preserved by the map.  If $\hodge$ denotes the Hodge duality operation
on $\Sigma$, our equations may be written as $\omegatil = \hodge
(T\omega)$ where we interpret $\omega$ and $\omegatil$ as pull backs
to $\Sigma$.  It is the desire to have the $\hodge$ operation that
introduces a $-1$ in (\ref{eq:minus}) even though in principle the
$-1$ could be absorbed\footnote{We may have to allow $T_{\pm} \in
\Orth(n)$.} into $T_{-}$.  Here we show that there are interesting
maps between ``waves'' on $M$ and ``waves'' on $\Mtil$ that preserve
natural geometrical structures.  This may be of interest to
researchers who study two dimensional wave equations.

This paper is organized in the following way.  In
Section~\ref{sec:pedagogic} we discuss pseudoduality between the
$2$-sphere $S^{2}$ and the $2$-dimensional hyperbolic space $H^{2}$ very
concretely.  In particular we explicitly verify the necessity for the
the orthogonal matrix valued function $T$.  In
Section~\ref{sec:constant} we become a bit more abstract but still
concrete by working with explicit metrics and show that there is
pseudoduality  between $S^{n}$ and $H^{n}$.  In
Section~\ref{sec:general} we assume general metrics and show that the
manifolds $M$ and $\Mtil$ must be symmetric spaces with the ``opposite
curvatures''.  In Section~\ref{sec:dualsym} we construct many examples
by discussing the theory of dual symmetric spaces. 
Section~\ref{sec:discussion} is a discussion of the results of this
article.

\section{A Pedagogic Example}
\label{sec:pedagogic}

It is worthwhile to be very concrete and to consider the pseudoduality
between strings moving on a $2$-sphere $S^{2}$ and those moving on a
$2$-hyperboloid $H^{2}$.  This example illustrates the necessity for
the matrix $T$.  We use a different coordinate version of the constant
curvature metric than in Section~\ref{sec:constant} to emphasize that
everything is independent of the choice of coordinates.  The
respective constant curvature metrics on $S^{2}$ and $H^{2}$ in polar
normal coordinates are
\begin{eqnarray*}
    ds^{2} & = & dr^{2} + \frac{\sin^{2}\alpha r}{\alpha^{2}}d\theta^{2}\;,
    \\
    d\tilde{s}^{2} & = & d\rtil^{2} + \frac{\sinh^{2}\alphatil
    \rtil}{\alphatil^{2}}d\thetatil^{2}\;.
\end{eqnarray*}
The equations of motion for the sigma models are:
\begin{eqnarray}
    \partial^{2}_{+-} r & = & (\sin \alpha r \cos\alpha r)/\alpha \;
    	\partial_{+}\theta \partial_{-}\theta\;,
	\label{eq:EOMr}\\
    (\sin\alpha r)/\alpha\; \partial^{2}_{+-} \theta & = & 
    	- \cos \alpha r (\partial_{+}r \partial_{-}\theta 
	+ \partial_{-}r \partial_{+}\theta)\;,
	\label{eq:EOMtheta}\\
    \partial^{2}_{+-} \rtil & = & 
    	(\sinh \alphatil \rtil \cosh\alphatil \rtil)/\alphatil \; 
    	\partial_{+}\thetatil \partial_{-}\thetatil\;,
	\label{eq:EOMrtil}\\
    (\sinh\alphatil \rtil)/\alphatil\; \partial^{2}_{+-} \thetatil & = & 
    	- \cosh \alphatil \rtil (\partial_{+}\rtil \partial_{-}\thetatil 
	+ \partial_{-}\rtil \partial_{+}\thetatil)\;.
	\label{eq:EOMthetatil}
\end{eqnarray}

This elementary example illustrates the importance of the matrix $T$. 
Assume $T=I$ then two of the duality equations in this coordinate
system would be $\partial_{+}\rtil = \partial_{+} r$ and
$\partial_{-}\rtil = -\partial_{-} r$.  This would imply that
$\partial^{2}_{+-}r=0$ and $\partial^{2}_{+-}\rtil=0$ but these are
not the equations of motion. We will see that with a $T\neq I$ we can 
construct a pseudoduality transformation.

First we need an orthonormal frame. On $S^{2}$ we have $\omega^{r} = dr$ and 
$\omega^{\theta}=\sin\alpha r/\alpha\;d\theta$. From this it 
follows that the connection and the curvature are respectively given by 
$\omega_{r\theta}= -\cos(\alpha r) \; d\theta$ and
$d\omega_{r\theta} = \alpha^{2}\omega^{r}\wedge\omega^{\theta}$.
On $H^{2}$ we have $\omegatil^{\rtil} = d\rtil$ and 
$\omegatil^{\thetatil}=\sinh\alpha \rtil/\alpha\;d\thetatil$.  From 
this we see that the connection and the curvature are respectively given by 
$\omegatil_{\rtil\thetatil}= -\cosh(\alphatil \rtil) \; d\thetatil$ and 
$d\omegatil_{\rtil\thetatil} = 
-\alphatil^{2}\omegatil^{\rtil}\wedge\omegatil^{\thetatil}$.

We need a $2\times 2$ orthogonal matrix that we parametrize by a 
rotation angle $\phi$. The duality equations are
\begin{equation}
    \left(
    \begin{array}{c}
        \partial_{\pm}\rtil  \\
        \sinh (\alphatil\rtil)/\alphatil\; \partial_{\pm}\thetatil
    \end{array}
    \right) = \pm
    \left(
    \begin{array}{rr}
        \cos\phi & \sin\phi  \\
        -\sin\phi & \cos\phi
    \end{array}
    \right)
    \left(
    \begin{array}{c}
        \partial_{\pm} r  \\
        \sin (\alpha r)/\alpha \; \partial_{\pm}\theta
    \end{array}
    \right)\,.
    \label{eq:dualrot}
\end{equation}
The integrability conditions on the above lead to the equation
\begin{equation}
	d\phi = -\cosh(\alphatil\rtil)\; d\thetatil
		+ \cos(\alpha r)\; d\theta\,.
	\label{eq:dphi}
\end{equation}
This is integrable if $\alpha=\alphatil$ where \emph{integrable} means 
integrable modulo the equations of motion.

What happens to the point particle geodesics on $S^{2}$?  One easily
verifies that for constant $a$, the functions $r=a\cdot
(\sigma^{+}+\sigma^{-}) = a\cdot (2\tau)$ and $\theta=0$ give a
solution of equations (\ref{eq:EOMr}) and (\ref{eq:EOMtheta}).  This
corresponds to ``particle geodesics'' on $S^{2}$.  We would like to
understand what are the dual solutions to the particle geodesics.  If
we note that $\partial_{\pm}r = a$ and $\partial_{\pm}\theta=0$ we see
that the duality equations become
\begin{eqnarray}
    \partial_{\pm}\rtil & = & \pm a \cos\phi
    \label{eq:dr}  \\
    \sinh (\alphatil\rtil)/\alphatil\; \partial_{\pm}\thetatil& = & 
    	\mp a \sin\phi
    \label{eq:dtheta}
\end{eqnarray}
An immediate consequence of the above is that
$(\partial_{+}+\partial_{-})\rtil=0$ and
$(\partial_{+}+\partial_{-})\thetatil=0$.  We conclude that
$\rtil=\rtil(\sigma^{+}-\sigma^{-}) =\rtil(2\sigma)$ and
$\thetatil=\thetatil(\sigma^{+}-\sigma^{-}) =\thetatil(2\sigma)$. 
Thus we get static solutions on the hyperboloid.  Note that
(\ref{eq:dphi}) immediately tells us that
$\phi=\phi(\sigma^{+}-\sigma^{-}) =\phi(2\sigma)$.  Since everything
is a function of $2\sigma =\sigma^{+}-\sigma^{-}$ the situation
reduces to functions of a single variable.  The transformed solutions
will be static solutions ($\tau$ independent).  Note that equations
(\ref{eq:dr}) and (\ref{eq:dtheta}) lead to
\begin{equation}
    (\partial_{+}\rtil)^{2} + 
    \left(\frac{\sinh\alphatil\rtil}{\alphatil}\; 
    \partial_{+}\thetatil\right)^{2} = a^{2}\,.
    \label{eq:energy}
\end{equation}
This is the ``conservation of energy'' equation associated with the 
particle lagrangian
$$
	L = \half\left[ (\partial_{+}\rtil)^{2} + 
	\left(\frac{\sinh\alphatil\rtil}{\alphatil}\; 
	\partial_{+}\thetatil\right)^{2} \right]
$$
This is a standard problem in classical mechanics. The canonical 
momentum
$$
	\tilde{J} = \frac{\partial
	L}{\partial(\partial_{+}\thetatil)} =
	\left(\frac{\sinh\alphatil\rtil}{\alphatil} \right)^{2}\;
	\partial_{+}\thetatil
$$
is a constant of the motion. Thus the energy integral 
(\ref{eq:energy}) may be written as
\begin{equation}
    (\partial_{+}\rtil)^{2} +
    \frac{\alphatil^{2}\tilde{J}^{2}}{\sinh^{2}\alphatil\rtil} =
    a^{2}\;
\end{equation}
and this problem is reducible to quadrature.  This is interesting
because we know that there exists crystallographic subgroups $\Gamma$
of $\SL(2,\bbR)$ such that $H/\Gamma$ is a genus $g>1$ compact Riemann
surface.  Such a surface has closed geodesics of minimal length and
some of these are the static ``soliton-like'' solutions 
$(\rtil(2\sigma),\thetatil(2\sigma))$ we are
constructing above.

\section{Constant Curvature Metric}
\label{sec:constant}

Before presenting the general theory we study the special case of
constant curvature spaces.  Again we take $T_{+}=T_{-}=T$. 
The two dimensional nonlinear sigma model on a space with constant
positive curvature $k>0$ is shown to be pseudodual to the nonlinear
sigma model on a space with constant negative curvature $-k$. 
The pseudoduality equations are used to map solutions of one model into
solutions of the other model.

Locally, a metric on a space with constant curvature is may be written 
in the Poincar\'{e} form
\begin{equation}
    ds^{2} = \frac{dx^{i}\otimes dx^{i}}{\left(1 + k x^{2}\right)^{2}}\;.
    \label{eq:constcurv}
\end{equation}
An orthonormal frame in this metric is
\begin{equation}
    \omega^{i} = \frac{dx^{i}}{1 + kx^{2}}\;.
    \label{eq:frame}
\end{equation}
The connection one-forms (with respect to the orthonormal frame) are
\begin{equation}
    \omega_{ij} = 2k\; \frac{x^{i}dx^{j}-x^{j}dx^{i}}{1+kx^{2}}\;.
    \label{eq:connection}
\end{equation}
The first Cartan structural equation is $d\omega^{i}= 
-\omega_{ij}\wedge\omega^{j}$. The second structural equation gives the
the curvature two forms
\begin{equation}
    \Omega_{ij} = d\omega_{ij} + \omega_{ik}\wedge\omega_{kj} 
    =\half R_{ijkl}\omega^{k}\wedge\omega^{l} = 
    4k\,\omega^{i}\wedge\omega^{j}\;.
    \label{eq:curvature}
\end{equation}
Technically the curvature is $4k$.

The lagrangian for the sigma model is
\begin{equation}
    \mathcal{L} = 
    \frac{2\dplus x^{i} \dminus x^{i}}{\left(1 + k x^{2}\right)^{2}}\;.
    \label{eq:lagrangian}
\end{equation}
The equations of motion associated with this lagrangian are
\begin{equation}
    \dalem x^{i} = 2k\;
    \frac{x^{j}\dplus x^{j}\dminus x^{i}
    	+ x^{j}\dminus x^{j} \dplus x^{i}
	- x^{i}\dplus x^{j} \dminus x^{j}}{1 + kx^{2}}\;.
    \label{eq:eom}
\end{equation}

Assume we have two constant curvature spaces $M$ and $\Mtil$ with
respective curvatures $k$ and $\ktil$.  We would like to see if it is
possible to have a duality transformation between them.  Assume we
have a solution $x(\sigma)$ of the sigma model on $M$.  We attempt to
construct a solution of the sigma model on $\Mtil$ by requiring that
\begin{eqnarray}
    \frac{\dplus \xtil^{i}}{1 + \ktil \xtil^{2}} & = & 
    	\Tij\; \frac{\dplus x^{j}}{1 + k x^{2}}\;,
    \label{eq:dualp}  \\
    \frac{\dminus \xtil^{i}}{1 + \ktil \xtil^{2}} & = & 
    	-\Tij\; \frac{\dminus x^{j}}{1 + k x^{2}}\;,
    \label{eq:dualm}
\end{eqnarray}
where $T$ is an orthogonal matrix, $\det T=1$.  Note that the stress
energy tensors satisfy $\Theta_{\pm\pm}=\widetilde{\Theta}_{\pm\pm}$. 

The first step towards the integrability conditions for the above 
system is to differentiate (\ref{eq:dualp}) with respect to 
$\sigma^{-}$ and 
(\ref{eq:dualm}) with respect to $\sigma^{+}$:
\begin{eqnarray}
    \frac{\dminus(\dplus \xtil^{i})}{1 + \ktil \xtil^{2}}
    	- 2\ktil \frac{\xtil^{j}\dminus \xtil^{j} \dplus 
    	\xtil^{i}}{(1 + \ktil \xtil^{2})^{2}} & = & 
	(\dminus \Tij) \frac{\dplus x^{j}}{1 + kx^{2}} 
    \nonumber  \\
     & + & \Tij\; \frac{\dalem x^{j}}{1 + k x^{2}} -
     2 k \Tij\; \frac{x^{k}\dminus x^{k}\dplus x^{j}}{(1 + k x^{2})^{2}}\;,
    \label{eq:intp} \\
    -\frac{\dplus(\dminus \xtil^{i})}{1 + \ktil \xtil^{2}}
    	+ 2\ktil \frac{\xtil^{j}\dplus \xtil^{j} \dminus 
    	\xtil^{i}}{(1 + \ktil \xtil^{2})^{2}} & = & 
	(\dplus \Tij) \frac{\dminus x^{j}}{1 + kx^{2}} 
    \nonumber  \\
     & + & \Tij\; \frac{\dalem x^{j}}{1 + k x^{2}} -
     2 k \Tij\; \frac{x^{k}\dplus x^{k}\dminus x^{j}}{(1 + k x^{2})^{2}}\;.
    \label{eq:intm}
\end{eqnarray}
By imposing the integrability conditions $\dplus(\dminus\xtil^{i}) = 
\dminus(\dplus\xtil^{i})$ we can add the above two equations to 
eliminate $\dalem \xtil$ and by imposing {\em only} the equations of motion 
for $x^{i}$ and the duality relations we obtain
\begin{eqnarray}
    (\dminus \Tij) \frac{\dplus x^{j}}{1 + kx^{2}}
    + (\dplus \Tij) \frac{\dminus x^{j}}{1 + kx^{2}}
     & = & 2\ktil\;
     	\frac{\xtil^{j}\dminus \xtil^{i}
	- \xtil^{i}\dminus \xtil^{j}}{1 + \ktil \xtil^{2}}\;
	\Tjk\; \frac{\dplus x^{k}}{1 + k x^{2}}
      \nonumber\\
     & + & 2\ktil\;
     	\frac{\xtil^{j}\dplus \xtil^{i}
	- \xtil^{i}\dplus \xtil^{j}}{1 + \ktil \xtil^{2}}\;
	\Tjk\; \frac{\dminus x^{k}}{1 + k x^{2}}
      \nonumber\\
     & - & 2k\;
     	\Tij\; \frac{x^{k} \dminus x^{j} - x^{j} \dminus x^{k}}{1 + kx^{2}}
	\; \frac{\dplus x^{k}}{1 + k x^{2}}
	\nonumber\\
     & - & 2k\;
     	\Tij\; \frac{x^{k} \dplus x^{j} - x^{j} \dplus x^{k}}{1 + kx^{2}}
	\; \frac{\dminus x^{k}}{1 + k x^{2}}\;.
    \label{eq:integrability}
\end{eqnarray}
The particular combinations in the above follow from the general
theory and the form of (\ref{eq:connection}).  Note that at any point
$\sigma$ we can make $\partial_{\pm}x^{i}(\sigma)$ arbitrary thus we
need that
\begin{eqnarray}
    \dminus \Tik & = &  2\ktil\; \frac{\xtil^{j}\dminus \xtil^{i}
	- \xtil^{i}\dminus \xtil^{j}}{1 + \ktil \xtil^{2}}\;\Tjk
      -  2k\;
     \Tij\; \frac{x^{k} \dminus x^{j} - x^{j} \dminus x^{k}}{1 + 
     kx^{2}}\;,
	\label{eq:dmT} \\
    \dplus \Tik & = & 2\ktil\;
     	\frac{\xtil^{j}\dplus \xtil^{i}
	- \xtil^{i}\dplus \xtil^{j}}{1 + \ktil \xtil^{2}}\;\Tjk
      -  2k\;\Tij\; 
	\frac{x^{k} \dplus x^{j} - x^{j} \dplus x^{k}}{1 + kx^{2}}\;.
	\label{eq:dpT}
\end{eqnarray}
We interpret the above as the pullback under the map $\sigma 
\to (x(\sigma), \xtil(\sigma))$ of the $1$-form
\begin{equation}
    d\Tij = -\omegatil_{ik}\Tkj + \Tik\omega_{kj}\;.
    \label{eq:dT}
\end{equation}

We can immediately verify that if one inserts (\ref{eq:dmT}) into 
equation~(\ref{eq:intp}) we get the equations for motion for 
$\xtil$, \emph{i.e.}, equations~(\ref{eq:eom}) with all quantities with 
tildes. Thus far we have shown that if one starts with a solution of 
the sigma model defined by lagrangian (\ref{eq:lagrangian}) and we 
impose the generalized duality equations then we find a solution to 
the ``tilde'' sigma model provided that we can satisfy equations 
(\ref{eq:dpT}) and (\ref{eq:dmT}).

The integrability conditions for the $dT$ equation are found by taking the 
exterior derivative leading to
\begin{equation}
    0 = - \widetilde{\Omega}_{ik}\Tkj + \Tik\Omega_{kj}
    \label{eq:dTintegrability}
\end{equation}
In our case the curvature is very simple and the above reduces to 
\begin{equation}
    -4\ktil\, \omegatil^{i}\wedge\omegatil^{k}\Tkj + 4k\, \Tik 
    \omega^{k}\wedge\omega^{j}=0\;.
    \label{eq:kktil}
\end{equation}

First we discuss what equation~(\ref{eq:dTintegrability}) \emph{is
not} and afterwards we discuss what \emph{it is}.  This equation shows
up in the study of local riemannian isometries \cite{BCG3}.  If we
have two riemannian manifolds $M$ and $\Mtil$ and a map $f:M\to\Mtil$
then the conditions that $f$ be a local isometry is that there exists
an orthogonal transformation $T$ such that
$\omegatil^{i}=\Tij\omega^{j}$.  When one works out the integrability
conditions for this system one finds that (\ref{eq:dT}) must be
satisfied along with its integrability condition which is
(\ref{eq:dTintegrability}).  This tells us that
$\widetilde{R}_{ijkl}T^{k}{}_{m}T^{l}{}_{n}= T^{i}{}_{k}T^{j}{}_{l}
R_{klmn}$.  There are further integrability conditions which tell us
that the covariant derivatives of the curvatures also satisfy similar
relations.  This is the classical theorem of Christoffel on local
isometries between riemannian manifolds as reformulated by E.~Cartan. 
In our constant curvature example the requirement simply becomes
$k=\ktil$.

Our situation is very different.  In effect our setup\footnote{We
avoid discussing jet bundles.} is a map from $\Sigma$ to
$M\times\Mtil$.  This tells us that
\begin{eqnarray*}
    \omega^{i}= \frac{(\partial_{a}x^{i})d\sigma^{a}}{1 + k x^{2}} 
    \quad\mbox{and}\quad
    \omegatil^{i} =\frac{(\partial_{a}\xtil^{i})d\sigma^{a}}{1 + 
    \ktil \xtil^{2}}\;.
\end{eqnarray*}
Substituting these into (\ref{eq:kktil}), using duality relations 
(\ref{eq:dualp}) and (\ref{eq:dualm}), and noting that at 
any $(\sigma^{+},\sigma^{-})$ we are free to arbitrarily specify 
$\partial_{\pm}x^{i}(\sigma)$ leads to $k = -\ktil$.  The 
change in sign is due to the negative sign in (\ref{eq:dualm}). Thus we 
discover that the duality equations can be implemented if the 
constant curvature manifolds have the opposite curvature.

\section{General Theory}
\label{sec:general}

This problem is best analyzed in the bundle of orthogonal
frames~\cite{Alvarez:psd2}.  Because most physicists are not familiar
with this approach we work things out on the base manifold
$M\times\Mtil$.  First thing to do is to take the exterior derivative
of (\ref{eq:special})
$$
    d\xtil_{\pm} = \pm (dT) x_{\pm} \pm T dx_{\pm}
$$
and use the definitions of the covariant derivative (\ref{eq:covder}) 
to obtain
$$
    -\omegatil \xtil_{\pm} + \xtil_{\pm a}d\sigma^{a}
    = \pm (dT) x_{\pm} \mp T \omega x_{\pm}
    \pm T x_{\pm a} d\sigma^{a}\,.
$$
If we use the duality equations (\ref{eq:special}) we have
$$
    \mp \omegatil T x_{\pm} + \xtil_{\pm a}d\sigma^{a}
    = \pm (dT) x_{\pm} \mp T \omega x_{\pm}
    \pm T x_{\pm a} d\sigma^{a}\,.
$$
A little algebra shows that
\begin{equation}
     \xtil_{\pm a}d\sigma^{a}
    = \pm (dT  -  T \omega  + \omegatil T)x_{\pm}
    \pm T x_{\pm a} d\sigma^{a}\,.   
    \label{eq:pseudodual}
\end{equation}
We wish to isolate the integrability conditions so wedge the above 
with $d\sigma^{\pm}$.
$$
     \xtil_{\pm \mp}d\sigma^{\mp} \wedge d\sigma^{\pm}
    = \pm (dT  -  T \omega  + \omegatil T)x_{\pm} \wedge d\sigma^{\pm}
    \pm T x_{\pm \mp} d\sigma^{\mp} \wedge d\sigma^{\pm}\,.   
$$
We have two equations
\begin{eqnarray*}
     \xtil_{+ -}d\sigma^{-} \wedge d\sigma^{+}
    &=& + (dT  -  T \omega  + \omegatil T)x_{+} \wedge d\sigma^{+}
    + T x_{+ -} d\sigma^{-} \wedge d\sigma^{+}\,,   \\
     \xtil_{- +}d\sigma^{+} \wedge d\sigma^{-}
    &=& - (dT  -  T \omega  + \omegatil T)x_{-} \wedge d\sigma^{-}
    - T x_{- +} d\sigma^{+} \wedge d\sigma^{-}\,.   
\end{eqnarray*}
In principle we wish that the integrability conditions
$\xtil_{+-}=\xtil_{-+}$ are satisfied if the equations of motion
$x_{+-}=0$ hold.  Subsequently we would like that this implies that
$\xtil_{+-}=0$.  We might as well substitute $x_{+-}=0$ and
$\xtil_{+-}=0$ into the equations above and find
$$
    0 = (dT - T \omega + \omegatil T)x_{\pm}\wedge
    d\sigma^{\pm}\,.
$$
Since $x^{i}_{\pm}$ may be  arbitrarily specified at any $\sigma$ we 
have
$$
    0 = (dT - T \omega + \omegatil T)\wedge d\sigma^{\pm}\,.    
$$
Since $x:\Sigma\to M$ and $\xtil:\Sigma\to\Mtil$ are maps of a two 
dimensional worldsheet we conclude that on the worldsheet, the 
covariant derivative of $T$ vanishes
\begin{equation}
    dT - T \omega + \omegatil T =0\,.    
    \label{eq:covT}
\end{equation}
The reason is that the covariant differential of $T$ is a $1$-form. 
All our objects arise from maps with domain $\Sigma$ so the covariant
differential of $T$ is a $1$-form on $\Sigma$.  You are pulling back
both $\omega$ and $\omegatil$ to $\Sigma$.  Note that a covariantly
constant tensor is determined its value at one point on the
worldsheet; the values elsewhere are determined by parallel transport. 
In order to construct such a $T$ we need to verify the integrability
conditions for the above.  These lead to important constraints on the
geometry of $M$ and $\Mtil$.  Taking the exterior derivative and using
the Cartan structural equations leads to
$$
    -T^{i}{}_{k}\Omega_{kj} + \widetilde{\Omega}_{ik}T^{k}{}_{j}=0
$$
with special case (\ref{eq:dTintegrability}). Expanding the above gives
$$
    -\half T^{i}{}_{k}R_{kjlm}\omega^{l}\wedge\omega^{m}
    + \Rtil_{iklm}T^{k}{}_{j} \omegatil^{l}\wedge\omegatil^{m}=0\,.
$$
If we now substitute (\ref{eq:defvel}) and use the pseudoduality 
equations (\ref{eq:special}) we see that
\begin{equation}
    T^{i}{}_{k}T^{j}{}_{l}R_{klmn}= 
    -\widetilde{R}_{ijkl}T^{k}{}_{m}T^{l}{}_{n}\;.
    \label{eq:pCurv}
\end{equation}
Thus we conclude that the manifolds $M$ and $\Mtil$ ``have the 
opposite curvature''. Next we take the exterior derivative of the 
above to look for further conditions. If the covariant differential of 
$R$ is defined by $D R_{ijkl} = R_{ijkl;m}\omega^{m}$ and similarly 
for $\Rtil$ we find
$$
    T^{i}{}_{k}T^{j}{}_{l}R_{klmn;p} \omega^{p} =
    -\widetilde{R}_{ijkl;p}T^{k}{}_{m}T^{l}{}_{n}\omegatil^{p}\,.
$$
If we now substitute (\ref{eq:defvel}) into the above and use the 
pseudoduality equations (\ref{eq:special}) we obtain two equations 
since $d\sigma^{+}$ and $d\sigma^{-}$ are independent:
\begin{eqnarray*}
    T^{i}{}_{k}T^{j}{}_{l}R_{klmn;q} & = & 
    -\widetilde{R}_{ijkl;p}T^{k}{}_{m}T^{l}{}_{n} T^{p}{}_{q}\,,\\
    T^{i}{}_{k}T^{j}{}_{l}R_{klmn;q} & = & 
    +\widetilde{R}_{ijkl;p}T^{k}{}_{m}T^{l}{}_{n} T^{p}{}_{q}\,.
\end{eqnarray*}
The solution of the above is immediate
\begin{equation}
    R_{klmn;q}=0\quad\mbox{and}\quad \widetilde{R}_{ijkl;p} =0\,.
    \label{eq:DR0}
\end{equation}
The manifolds $M$ and $\Mtil$ must be \emph{locally symmetric spaces} with the 
``opposite curvature''. 

\section{Dual Symmetric Spaces}
\label{sec:dualsym}

There is a large class of examples of pairs of symmetric spaces with
opposite curvature.  These pairs are called dual symmetric
spaces.  The simplest pair is the $n$-sphere $S^{n}$ and the
$n$-dimensional hyperbolic space $H^{n}$.  More complicated pairs are
given by the Grassmann manifolds\footnote{$\Orth_{0}(p,q)$ is the
component of $\Orth(p,q)$ connected to the identity.}
$\SOrth(p+q)/\SOrth(p)\times\SOrth(q)$ and
$\Orth_{0}(p,q)/\SOrth(p)\times\SOrth(q)$.

The general theory of symmetric spaces is very extensive
\cite{Helgason:DG,Wolf:CC}.  For our purposes we take a lighter
approach \cite[Chapter 11]{ONeill:SRG} and a more restrictive view and
consider what are called \emph{normal symmetric spaces}.  These are
specified by a triplet of data $(G/H,\sigma,Q)$ where $G$ is a real
Lie group, $H$ is a closed subgroup of $G$ and $\sigma$ is an
involutive automorphism of $G$.  Let $\lieg$ be the Lie algebra of $G$
and denote the action of the automorphism $\sigma$ on $\lieg$ by $s$. 
Let $\lieh$ be the $+1$ eigenspace of $s$ and let $\liem$ be the $-1$
eigenspace of $s$.  Since $\half(I \pm s)$ are projectors we have
$\lieg=\lieh \oplus \liem$.  The following are also required in a 
normal symmetric space:
\begin{enumerate}
    \item If $F$ is the fixed point set of $\sigma$ and $F_{0}$ is its
    identity component then $F_{0}\subset H\subset F$.  This is a
    technical requirement that for our purposes we take it to mean
    that the Lie algebra of $H$ is $\lieh$.

    \item $Q$ is an $\Ad(G)$-invariant inner product on $\lieg$.  An
    ordinary symmetric space only requires $Q$ to be an
    $\Ad(H)$-invariant inner product\footnote{An $\Ad(H)$-invariant
    inner product on $\liem$ leads to a $G$-invariant metric on $M$, 
    see \cite[p. 312]{ONeill:SRG}.}
    on $\liem$.  Here $\Ad(G)$ is the adjoint action of the group $G$
    on its Lie algebra $\lieg$.
    
    \item $Q$ is $s$-invariant. This is not required for an 
    ordinary symmetric space.
\end{enumerate}
It follows that $[\lieh,\lieh]\subset\lieh$,
$[\lieh,\liem]\subset\liem$ and $[\liem,\liem]\subset\lieh$.  The
$s$-invariance of $Q$ tells us that the direct sum decomposition is an
orthogonal decomposition.  Note that $\lieh$ and $\liem$ are
orthogonal with respect to any $s$-invariant quadratic form such as
the Killing form $\Tr(\ad(X)\ad(Y))$.

We pick an origin for the symmetric space $G/H$ and associate $\liem$
with the tangent space at that point.  For a normal symmetric space
the sectional curvature associated to the $2$-plane spanned by $X,Y
\in \liem$ is given by the following simple formula~\cite[p. 319]{ONeill:SRG}
\begin{equation}
    K(X,Y) = 
    \frac{Q([X,Y],[X,Y])}{Q(X,X)Q(Y,Y)-Q(X,Y)^{2}}
    \label{eq:sectional}
\end{equation}
that requires the $\Ad(G)$-invariance of $Q$ in its derivation.
We remind the reader that knowing all sectional curvatures is
equivalent to knowing the curvature tensor \cite{ONeill:SRG} and that
they are related by
$$
    K(X,Y) = 
    \frac{R_{ijkl}X^{i}Y^{j}X^{k}Y^{l}}{Q(X,X)Q(Y,Y)-Q(X,Y)^{2}}\;.
$$

Normal symmetric spaces $(G/H,\sigma,Q)$ and 
$(\Gtil/\Htil,\sigtil,\Qtil)$ are said to be \emph{dual 
symmetric spaces} if there exist
\begin{enumerate}
    \item a Lie algebra isomorphism $S:\lieh\to
    \tilde{\lieh}$ such that $\Qtil(S V,S W) = - Q(V,W)$ 
    for all $V,W \in \mathfrak{h}$.

    \item a linear isometry $T:\liem \to \widetilde{\liem}$ such that 
    $[TX,TY]=-S[X,Y]$ for all $X,Y \in \liem$.
\end{enumerate}
In item (1) above the Lie algebra isomorphism tells us that brackets
in $\lieh$ are the same as in $\tilde{\lieh}$.  While the isometry in
item (2) tells us that inner product on $\liem$ is the same as in
$\widetilde{\liem}$.

For dual symmetric spaces it is easy to see that the 
sectional curvatures are related by $\widetilde{K}(TX,TY) = -K(X,Y)$.

It is worthwhile to work this out explicitly for the example of the dual
symmetric spaces $\SOrth(p+q)/\SOrth(p)\times\SOrth(q)$ and
$\Orth_{0}(p,q)/\SOrth(p)\times\SOrth(q)$.  For $g\in G=\SOrth(p+q)$
or $g\in \Gtil=\Orth_{0}(p,q)$ we take $\sigma$ and
$\sigtil$ to be given by
$$
    g \mapsto
    \left(
    \begin{array}{cc}
        I_{p} & 0  \\
        0 & -I_{q}
    \end{array}
    \right) g
    \left(
    \begin{array}{cc}
        I_{p} & 0  \\
        0 & -I_{q}
    \end{array}
    \right)\,.
$$
One easily verifies that $\lieh=\tilde{\lieh} = \so(p) \oplus \so(q)$ 
where the matrices are of the form
$$
    \left(
    \begin{array}{cc}
        a & 0  \\
        0 & c
    \end{array}
    \right) \,
$$
where $a \in \so(p)$ and $c\in \so(q)$. One also sees that 
$X\in\liem$ and $\widetilde{X}\in \widetilde{\liem}$ are of the form
$$
    X = \left(
    \begin{array}{cc}
        0 & -b^{t}  \\
        b & 0
    \end{array}
    \right)
    \quad\mbox{and}\quad
    \widetilde{X} = \left(
    \begin{array}{cc}
        0 & b^{t}  \\
        b & 0
    \end{array}
    \right)\,,
$$
where $b$ is an arbitrary $q\times p$ matrix.  For the inner products
we take $Q(X,Y) = -\half\Tr(XY)$ and
$\Qtil(\widetilde{X},\widetilde{Y}) =
+\half\Tr(\widetilde{X}\widetilde{Y})$.  These will be ``riemannian''
dual symmetric spaces because the metrics on $\liem$ and
$\widetilde{\liem}$ are positive definite.  We take the subgroups $H$
and $\Htil$ to be $\SOrth(p)\times\SOrth(q)$ and the Lie algebra
isomorphism $S:\lieh \to \tilde{\lieh}$ is the identity
transformation.  The isometry $T:\liem\to \widetilde{\liem}$ is taken
to be
$$
    T:
    \left(
    \begin{array}{cc}
        0 & -b^{t}  \\
        b & 0
    \end{array}
    \right)
    \mapsto
    \left(
    \begin{array}{cc}
        0 & b^{t}  \\
        b & 0
    \end{array}
    \right)\,.
$$    
A brief computation shows that $[TX,TY] = -[X,Y]$.  From this we see
that $\SOrth(p+q)/\SOrth(p)\times\SOrth(q)$ is a space with positive
sectional curvature and $\Orth_{0}(p,q)/\SOrth(p)\times\SOrth(q)$ is a
space with negative sectional curvature.

A more extensive discussion of dual symmetric spaces requires a 
thorough discussion of the theory of orthogonal involutive Lie 
algebras (orthogonal symmetric Lie algebras), 
see~\cite{Helgason:DG,Wolf:CC}.

\section{Ivanov's Construction}
\label{sec:ivanov}

In this article we have limited ourselves to studying sigma models on
an arbitrary riemannian manifold with vanishing $3$-form $H_{ijk}$. 
Ivanov analyzed  a subset of those models.  He studied sigma models
that are Lie group theoretic in origin.  In this case we can use
special properties of Lie groups to simplify the analysis.  Assume we
have a normal symmetric space $G/H$ as in Section~\ref{sec:dualsym}. 
We can choose an orthonormal basis that respects the decomposition
$\lieg = \lieh \oplus \liem$.  If $\{T_{i}\}$ is such a basis with Lie
bracket relations
$$
    [T_{i}, T_{j}] = f^{k}{}_{ij}T_{k}
$$
then the $\Ad(G)$-invariance of $Q$ tells us that $f_{ijk}$ are 
totally antisymmetric. The indices $a,b,c,\ldots$ are associated to  
the basis elements that are in  $\lieh$ and the indices 
$\alpha,\beta,\gamma,\ldots$ are associated to basis elements in 
$\liem$. With this notation the Lie algebra bracket relations are
\begin{eqnarray}
    [T_{a},T_{b}] & = & f^{c}{}_{ab}T_{c}\,,
    \label{eq:hhh}  \\ \relax
    [T_{a},T_{\beta}] & = & f^{\gamma}{}_{a\beta}T_{\gamma}\,,
    \label{eq:hmm}  \\ \relax
    [T_{\alpha},T_{\beta}] & = & f^{c}{}_{\alpha\beta}T_{c}\,.
    \label{eq:mmh}
\end{eqnarray}
Let $\theta^{i}$ be the the Maurer-Cartan forms for the Lie group $G$ 
that satisfy the Maurer-Cartan equations
$$
    d\theta^{i} + \half f^{i}{}_{jk}\theta^{j}\wedge\theta^{k}
    =0\;.
$$
Using the decomposition $\lieg=\lieh\oplus\liem$ we may write the 
equations above as
\begin{eqnarray}
    d\theta^{a} + \half f^{a}{}_{bc}\theta^{b}\wedge \theta^{c}
    & = & - \half f^{a}{}_{\beta\gamma} 
    \theta^{\beta}\wedge\theta^{\gamma}\,,
    \label{eq:Hcurv}  \\
    d\theta^{\beta} + f^{\beta}{}_{b\gamma} 
    \theta^{b}\wedge\theta^{\gamma} & = & 0\;.
    \label{eq:covm}
\end{eqnarray}
It is worthwhile discussing the geometric meaning of the above
equations.  It is well known that $G$ is a principal $H$-bundle over
$G/H$.  The $H$-connection $\theta^{a}T_{a}$ on $G$ defines the
horizontal tangent spaces.  Its curvature is given by the right hand
side of (\ref{eq:Hcurv}).  Equation (\ref{eq:covm}) states that the
covariant differential of the $\liem$-valued $1$-form
$\theta^{\beta}T_{\beta}$ is zero.

The equations of motion nonlinear sigma model with target space $G/H$
may be formulated in terms of a map $g:\Sigma \to G$ that satisfies the
equations
\begin{eqnarray}
    d\theta^{a} + \half f^{a}{}_{bc}\theta^{b}\wedge \theta^{c}
    & = & - \half f^{a}{}_{\beta\gamma} 
    \theta^{\beta}\wedge\theta^{\gamma}\,,
    \label{eq:Hcurv1}  \\
    d\theta^{\beta} + f^{\beta}{}_{b\gamma} 
    \theta^{b}\wedge\theta^{\gamma} & = & 0\;, \\
    \label{eq:covm1}
    d(\hodge\theta^{\beta}) + f^{\beta}{}_{b\gamma} 
    \theta^{b}\wedge (\hodge\theta^{\gamma}) & = & 0\;.
    \label{eq:wave1} 
\end{eqnarray}
Here we implicitly interpret $\theta^{i}$ as the pullback under 
$g:\Sigma\to G/H$. More properly we should have 
written $g^{*}\theta^{i}$. 
We use $\hodge$ to denote the Hodge duality operation on $\Sigma$. On 
$1$-forms it is given by $\hodge(d\sigma^{\pm})= \pm d\sigma^{\pm}$. 
The first two equations above are just the pullbacks to $\Sigma$ of the 
Maurer-Cartan equations for $G$. The third equation (\ref{eq:wave1}) 
is essentially the wave equation.
For future reference we note that if $\alpha$, $\beta$ are $1$-forms on 
$\Sigma$ then 
\begin{equation}
    (\hodge \alpha)\wedge(\hodge\beta) = - 
    \alpha\wedge\beta\,,
    \label{eq:hodge}
\end{equation}
and that $(\hodge)^{2}\alpha=\alpha$.

Ivanov observed that if we define $\thetatil^{\beta} = 
\hodge\theta^{\beta}$ and $\thetatil^{a}=\theta^{a}$ then the 
equations above become
\begin{eqnarray}
    d\thetatil^{a} + \half f^{a}{}_{bc}\thetatil^{b}\wedge \thetatil^{c}
    & = & + \half f^{a}{}_{\beta\gamma} 
    \thetatil^{\beta}\wedge\thetatil^{\gamma}\,,
    \label{eq:Hcurv2}  \\
    d\thetatil^{\beta} + f^{\beta}{}_{b\gamma} 
    \thetatil^{b}\wedge\thetatil^{\gamma} & = & 0\;, \\
    \label{eq:covm2}
    d(\hodge\thetatil^{\beta}) + f^{\beta}{}_{b\gamma} 
    \thetatil^{b}\wedge (\hodge\thetatil^{\gamma}) & = & 0\;.
    \label{eq:wave2} 
\end{eqnarray}
These equations may be interpreted as the equations of motion for a
sigma model on the symmetric space $\Gtil/H$ where $\Gtil$ is a real
Lie group with real Lie algebra $\tilde{\lieg} = \lieh \oplus
(i\liem)$.  This Lie algebra has Lie brackets given by
\begin{eqnarray}
    [\Ttil_{a},\Ttil_{b}] & = & f^{c}{}_{ab}\Ttil_{c}\,,
    \label{eq:hhhtil}  \\ \relax
    [\Ttil_{a},\Ttil_{\beta}] & = & f^{\gamma}{}_{a\beta}\Ttil_{\gamma}\,,
    \label{eq:hmmtil}  \\ \relax
    [\Ttil_{\alpha},\Ttil_{\beta}] & = & -f^{c}{}_{\alpha\beta}\Ttil_{c}\,.
    \label{eq:mmhtil}
\end{eqnarray}
In the above we have $\Ttil_{a}=T_{a}$ and $\Ttil_{\beta}=
iT_{\beta}$.  Note that if you think of $\Gtil$ as a principal
$H$-bundle then the curvature of $\Gtil$, given by (\ref{eq:Hcurv2}), is
``opposite'' to that of $G$, given by (\ref{eq:Hcurv1}).

\section{Discussion}
\label{sec:discussion}

In \cite{Alvarez:2000bh} it was shown that a necessary condition for
target space duality is that the target spaces be parallelizable
manifolds.  In the scenario presented here where we only require ``on
shell'' pseudoduality and we see that the parallelizable requirement
is weakened substantially but we still find natural geometric
restrictions on the target spaces.  In the models considered here
where the $3$-forms $H$ and $\Htil$ vanish we saw that pseudoduality
requires that the target spaces be symmetric spaces with the
``opposite curvature'' generalizing a result of Ivanov
\cite{Ivanov:1987yv}.  The class of riemannian dual symmetric spaces
provides a wealth of examples.  In particular we studied explicitly
the example of dual Grassmann manifolds
$\SOrth(p+q)/\SOrth(p)\times\SOrth(q)$ and
$\Orth_{0}(p+q)/\SOrth(p)\times\SOrth(q)$.  We also saw the importance
of the isometry $T$.  Note that we never explicitly solved for $T$ but
we discovered conditions that guarantee its existence.  Finally it
should be emphasized that $T$ depends on the path since it comes from
integrating (\ref{eq:covT}) along the path.

We note that equations (\ref{eq:special}) may also be written as
\begin{eqnarray}
    \xtil_{\tau} & = & T x_{\sigma}\,,
    \label{eq:psd00}  \\
    \xtil_{\sigma} & = & T x_{\tau}\,.
    \label{eq:psd11}
\end{eqnarray}
Thus we immediately see that ``particle-like'' solutions ($\sigma$ 
independent) on $M$ get mapped into static ``soliton-like'' solutions 
on $\Mtil$ and vice-versa. If say $\Mtil$ has noncontractible loops 
then there will be stable ``soliton-like'' solutions.

Recently Evans and Mountain \cite{Evans:2000qx} constructed an
infinite number of local conserved commuting charges on sigma models
with the target space being a compact symmetric space.  It would be
interesting to apply the results of this paper to the construction of
Evans and Mountain.

\section*{Acknowledgments}

I would like to thank T.L.~Curtright, H.~Fenderya, L.A.~Ferreira,
L.~Mezincescu, R.~Nepomechie, J.~S\'{a}nchez Guill\'{e}n, I.M.~Singer
and C.~Zachos for comments and discussions.  I would also thank
E.~Ivanov for bringing to my attention reference~\cite{Ivanov:1987yv}. 
This work was supported in part by National Science Foundation grant
PHY--9870101.

\providecommand{\href}[2]{#2}\begingroup\raggedright\endgroup

\end{document}